# The Interplay between Human and Machine Agency

J. Brian Pickering, Vegard Engen and Paul Walland

IT Innovation Centre
Gamma House, Enterprise Road
Southampton, SO16 7NS, UK
`{jbp,ve,pww}@it-innovation.soton.ac.uk`

**Abstract.** Human-machine networks affect many aspects of our lives: from sharing experiences with family and friends, knowledge creation and distance learning, and managing utility bills or providing feedback on retail items, to more specialised networks providing decision support to human operators and the delivery of health care via a network of clinicians, family, friends, and both physical and virtual social robots. Such networks rely on increasingly sophisticated machine algorithms, e.g., to recommend friends or purchases, to track our online activities in order to optimise the services available, and assessing risk to help maintain or even enhance people's health. Users are being offered ever increasing power and reach through these networks by machines which have to support and allow users to be able to achieve goals such as maintaining contact, making better decisions, and monitoring their health. As such, this comes down to a synergy between human and machine agency in which one is dependent in complex ways on the other. With that agency questions arise about trust, risk and regulation, as well as social influence and potential for computer-mediated self-efficacy. In this paper, we explore these constructs and their relationships and present a model based on review of the literature which seeks to identify the various dependencies between them.

**Keywords:** General: HCI methods and theories · Human-machine networks · Agency · Trust · Modelling · Self-efficacy

## 1 Introduction

A definition of agency based on the notion of non-deterministic behaviours [1] fails to recognise the increasing variety and complexity of human-machine networks [1] (HMNs) [2], the intention of technology designers [3], and active intervention by bots within social networks [4, 5]. The concept of agency is particularly problematic in human-machine interactions [6]. Machine or material agency may be seen as automation, which originally required some tolerance from human agents [7]. But this is no longer true: technology can actively support human activity [8], and manifests increasingly complex interaction types [9]. Machine and human agency may not be the same and yet equally valid [10]; machine agency may be just "perceived autonomy" [11]; and it certainly enables human agency [12]. Indeed, agency may well be becoming a social and group construct where both humans and machines play a part [13, 14]; and used effectively, agency may even lead to innovative review of working practice [15].

---

[1] In the following we use *human-machine network* and *network* interchangeably.

The enabling contribution of machine agents within a network may have an effect on self-efficacy. Bandura's original definition of self-efficacy as an individual's belief in their ability to be able to achieve a given objective [16–18] has also been applied to technology [19, 20] and its acceptance [21]. There are, however, constraints on the support and positive contribution of technology to human self-efficacy, not least in terms of anxiety and suspicion around technology use [22, 23]. This may be further exacerbated by increasing machine animism: it may not always be obvious what machines are doing or what information they are collecting on other agents in the network [24, 25]. With this in mind, regulation is seeking to impose safeguards [26], but with only limited success [27]. Any consequent perception of risk can undermine a willingness to engage in some online activities [28]. However, assuming scepticism can be kept low, regulation can reduce the negative effects of risk across a number of contexts, leading to an increased level of trust [29].

Bringing together some of these constructs, this paper proposes updates to a recent model of trust in information technology [30, 31] with a detailed exploration of self-efficacy as it relates to agency [3, 18]. Since human-machine networks can be characterised by independently varying levels of agency [2], either or both may influence self-efficacy [18]. Further, introducing regulation and perceived risk [32], how these influence agency and, in turn, behaviours within an HMN, need consideration [33, 34]. On the other hand, for this exploratory study, we do not consider affect [35] or other motivators such as social identity or task [36] which may mediate behaviour in online networks. Similarly, we discount environmental trust as factors affecting both behaviour and self-efficacy [37].

In the original study by Thatcher et al. [31], the final set of constructs was based on an extensive literature review [30], and its validation via an opportunity sample of students and IT professionals to establish inter-construct dependencies and correlations. The intent to explore technology was seen to be dependent on social norms and computer self-efficacy, while the influence of trust in IT and in support personnel is mediated by the technology acceptance model (TAM) constructs of perceived usefulness and ease of use. The relationship between interpersonal, organisational and technology trust is well motivated in light of theoretical considerations of trust transfer [38], and of trust as an overall organising factor [39]. However, given the various interactions between agents within an HMN, both human-to-human and machine-to-machine, the question remains whether McKnight et al. [30] had captured all relevant constructs. Further, participants were drawn from a narrow field who may show a priori increased propensity to engage and persevere with technology [40]. Revisiting and extending their original model with specific reference to HMNs, therefore, implies careful consideration of those constructs as well as participant selection in validating the resulting research model.

## 2 Modelling Trust

Interpersonal and person-to-organisation trust is based on the judgement of perceived benevolence, integrity and competence [41, 42]. More recently applied to technology and its acceptance, self-efficacy and agency in HMNs interact with one another as well as influence trust. It is, therefore, appropriate to reconsider a model of trust in online interactions.

### 2.1 Related Research

In a series of studies McKnight, Thatcher and their colleagues explored different constructs associated with trust in technology [30, 31]. They found that trust in technology and a willingness to explore its use could be predicted from individual propensities to trust. Further, context-specific factors including social context and an individual's computer self-efficacy were found to relate directly to this willingness to explore technology [31].

Other studies, however, highlight a range of different constructs. For example, an extensive literature review suggests personal, organisational as well as cultural factors as instrumental for online trust [43]. Many studies stress the social [44, 45], not least the importance of communication and group adherence [46] and self-efficacy [37]. However, this social dimension is closely connected with agency [47]. Still others explore the interplay of risk, assurances of privacy and security [29, 48]. Drawing all of this together suggests an extension to the Thatcher et al model to incorporate greater focus on the social on the one hand, but also regulation and risk perception on the other.

### 2.2 Research Model

The focus of the proposed research model is human behaviour in an HMN. For an HMN to be successful, and both gain and sustain participation, it needs to enable benefits for the human agents within the network. Although there are a number of human participants taking different roles, it may well be that the network benefits one group in a different, or preferential, way to others. This does not alter the fact that the network, and the machine agents within it, are established in order to provide benefit to the human actors in that network.

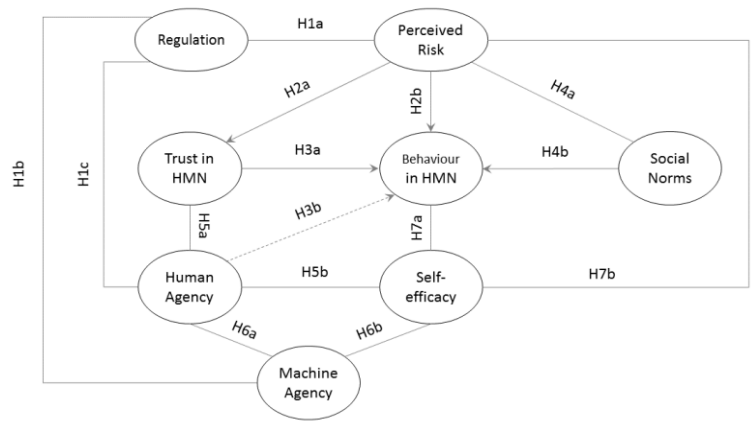

**Fig. 1.** Research Model

We are concentrating on human behaviour in the HMN, rather than human actions. This is an important distinction, since, as has been discussed elsewhere [49], human actors exhibit 'conscious intentionality', which is to say that the human actors have certain freedom of choice regarding their actions, whereas the machine agents exhibit 'programmed intentionality' in the sense that they can also influence behaviour, but do so according to their pre-determined programming and rule sets [50]. How a human behaves in a particular network and circumstance will depend upon a number of factors acting upon them, which are reflected in **Fig. 1**. The main proposition is that a human agent will behave in a manner which is determined by two considerations: their belief that they can achieve what they want in the network (self-efficacy) [16–18] and the level of risk that they perceive in performing those actions (trust) [28, 29]. What other constructs should be included is the purpose of this section.

First, we introduce the effect of regulation on agency, since regulation, whether legislative, standardisation or commercial restriction, will set limits on both machine agency and human agency [32]. Another effect of regulation is to modify the perceived risk involved in performing actions, positively or negatively [26, 32]. We identify perceived risk rather than actual or absolute risk, because it is perceived risk that determines human behaviour, not the actual level of risk, as established in a broad spectrum of research, including inter alia online consumer behaviour [51, 52], health care [53], and engineering and natural disasters [54]. For example, vanishingly rare events such as the murder of a child affect parents' perception of risk to their child, even though the actual risk is very small, and lower than the risk of injury at home. Thus, anything that modifies perceived risk is important, even if the actual level of risk is unchanged. We can, therefore, also conclude that trust is a reflection of perceived risk, and that behaviour is a reflection of willingness to accept a certain level of risk, both of which are based on belief, not on absolute or measurable parameters [31, 55]. Risk perception, trust and behaviour can all change with time, and can be modified by changes in circumstance, such as external influences (social norms) [56, 57].

Behaviour is also determined by ease of use of the network, expressed as ability to achieve pre-determined goals (self-efficacy). Machine agency can operate in a supportive role, enabling ease of use and hence supporting self-efficacy [50], leading to more positive or interactive behaviour and better achievement of objectives by the human agent. Limitations to the supportive ability of the machine agent, either through regulatory limitations or functional limitations, may reduce the self-efficacy they can support, leading to more cautious behaviour [31]. This would suggest an association with perceived risk: belief in one's own capabilities and therefore the ability to manage perceived risks.

## 3 Model Constructs and Research Hypotheses

As the central construct of our model, Human Behaviour in HMN[2] is affected directly or indirectly by all of the remaining constructs. In consequence, we have developed hypotheses for all the remaining constructs in the following sections. These will determine relationships with behaviour in HMN.

### 3.1 Regulation

We include laws (legislations), codes of practice and standards as part of this construct. As discussed above, regulations may constrain agency. It may also have enabling effects, such as technology standards improving on technology interoperability and ensure security and privacy for end-users [58], which are all key to HMNs dealing with personal data. Standardisation efforts, such as HL7 in healthcare[3], may, therefore, have a positive impact on both human and machine, as well as reducing the perceived risk. Miltgen and Smith [32] have already shown that higher levels of perceived regulatory protection is associated with a decrease in the risks that people perceive to their privacy. In this vein, we hypothesise the following.

**H1a**: Perceived risk is negatively correlated with changes in regulation.

However, regulation may also stifle innovation and constrain what actors are allowed to do, directly reducing both human and machine agency[4]. Reasons for regulations to be constrictive may be to, e.g., address concerns regarding the increasing autonomy of machines [2], raising ethical issues about responsibilities and accountability [59]. Thus, we hypothesise:

**H1b**: Machine agency is negatively correlated with increasing levels of regulation.

---

[2] Abbreviated to "Behaviour in HMN" in the remaining discussion
[3] HL7 – Health Level Seven. http://www.hl7.org.uk/
[4] EC H2020 SHiELD Project, 2017, http://cordis.europa.eu/project/rcn/207185_en.html

In the discussion above, we can see that the effect on human agency can vary, depending on the increase in regulations. On the one hand, it may increase human agency due to the reduction in the perceived risk (via trust in the HMN, as discussed further below). On the other hand, it may decrease human agency due to the reduction in machine agency, as hypothesised above. Therefore, we simply put forth the following hypothesis:

**H1c**: Regulation affects human agency.

### 3.2 Perceived Risk

The perception of risk and uncertainty was considered by Thatcher et al. [31], but was omitted from their proposed model. Here, we focus on the perception of risk as experienced by the human participants in an HMN. However, we do note that this also affects other actors, such as providers and even machines although their response is deterministic. While exploring the different types of risk that could affect trust and the behaviour of participants in an HMN is outside the scope of this paper, this could include any factors of an HMN pertaining to, e.g., monetary loss or loss of privacy, as discussed in [32, 60]. We focus here on the perception of risk experienced when participating in HMNs, which may well differ from actual risk as discussed above; the former affecting behaviour [31, 55], which pertains to the central construct we are interested in here (Behaviour in HMN). Pavlou [60], in proposing extensions to the TAM, found that trust influences perceived risk. Here we focus on how perceived risk affects trust in the HMN, leading to the following hypothesis:

**H2a**: The perception of risk negatively affects trust in an HMN.

While previous work, as discussed above [55, 60, 61], establishes that trust affects the perception of risk, we also hypothesise a direct relationship between Perceived Risk and Behaviour in HMN.

**H2b**: The perception of risk negatively affects the behaviour in an HMN.

### 3.3 Trust in HMNs

As a construct, trust pervades almost all interactions between individuals and is traditionally the result of perceived benevolence, competence and integrity [41]. Over time, it may be lost but also rebuilt [42], largely due to context and reassessment of behaviours and intention. If trust is an overall organising principle [39], then it makes sense to attempt to extend the construct to technology [30] and online networked activities [38]. So trust will affect self-efficacy [31, 37] and be associated with social norms [31]. It may well be that interpersonal and technology trust differ in the detail, but collectively influence a willingness to adopt the technology [62]. Once online, traditional behaviours occur: communication is important [46], and social trust will affect willingness to engage [63]. Further, increasing familiarity with the technology may well influence trust and thereby agency [20, 30], though there will be a moderating effect in relation to security risk or privacy exposure [48]. We therefore hypothesise that:

**H3a**: HMN trust will positively influence the behaviour in an HMN.

further

**H3b**: HMN trust will mediate human agency and the behaviour in an HMN.

### 3.4 Social norms

A significant research body on social norms in the offline world is gradually being extended into the virtual world, already finding parallels even down to eye gaze and social gestures [64] and group composition [65]. A willingness to engage online involves social pressures from the immediate social group, social identity [66] and trust in other network members [63]. Indeed the desire to be seen online will often motivate the adoption of strategies to mitigate against potential risk [56, 57], or even adapt structures such as reputation and social presence in decisions to engage, for instance in eCommerce [67]. Along with (computer) self-efficacy, social norms may also influence participation in social networks [68], and even encourage emergent and shared agency within the virtual group [13]. Thatcher et al. [31] were clearly right to include social norms in their model, though they did not necessarily explore the full implications of its influence. Specifically, social norms can reduce perceived risk, as well as directly encourage online engagement and behaviours. In consequence, we propose that:

**H4a**: Social norms will affect the extent to which perceived risk influences the behaviour in an HMN.

and

**H4b**: Social norms will directly and positively correlate with the behaviour in an HMN.

### 3.5 Human agency

A pragmatic definition of Human and Machine Agency in HMNs has been discussed in [3] on the basis of a review of social psychology literature, such as Structuration Theory [61] and Social Cognitive Theory [49]. Adopting the definition of agency from [3] we understand agency "as the capacity to perform activities in a particular environment in line with a set of goals/objectives that influence and shape the extent and nature of their participation". In practice, agency, therefore, indicates what a human actor can actually do in the network, and how this aligns with the objectives they would have for using the network, or their belief in their ability to achieve their goals. This, in turn, influences their behaviour in the network. We hypothesise a direct relationship between human agency and self-efficacy in terms of the behaviour in the HMN, as follows:

**H5a**: Trust in the HMN is positively correlated with human agency.
**H5b**: Human agency is related to self-efficacy.

### 3.6 Machine agency

As per [3], we can apply the same definition for Human Agency, as discussed above, to Machine Agency. While there are distinctions between the two, such as the lack of intentionality in machines [69, 70], they are increasingly active and visible participants in HMNs, even exhibiting human-like characteristics, capable of exerting influence and enhancing Human Agency [3]. The latter is due to a characteristic that Bandura [49] refers to as proxy agency, in which an agent may increase their own agency by utilising the capabilities of other agents, which could indeed be machines. However, it is far from clear whether machine agency might be perceived as a constraint on human activity itself or in overall processes [15]. Similarly a unidirectional relationship may not hold: human agency may well constrain machine agency if this means that human agents simply do not need the full capabilities of machines. Indeed, Følstad et al. [59] indicate a bi-directional and synergistic relationship, which warrants further exploration. Thus, we pose the following generic hypothesis:

**H6a**: Machine agency in an HMN is directly related to human agency.

The nature of this relationship may need more careful consideration. A similar issue arises regarding the relationship between Machine Agency and Self-efficacy, confounded by factors such as the age and cultural background of those engaging in the HMN. Whilst increasing machine agency may indeed increase the self-efficacy of certain population groups, it may have the opposite effect on others depending on their appraisal of technology [71].

**H6b**: Machine agency affects computer self-efficacy.

### 3.7 Computer self-efficacy

As stated, self-efficacy is a personal belief in one's ability to achieve [16, 18]; and in terms of technology use, often referred to as computer self-efficacy, it may be understood as internal (a belief that I can do it myself) or external (a belief that I can do it with appropriate support) [19]. It is assumed that younger people are more willing to engage with technology and see what happens, which seems to be the case [22]. Further, since on the one hand people change in their experience and expectations, and on the other technology develops, so we need to be sensitive to such change especially in our metrics [20]. In HMN terms, it turns out that self-efficacy is related to trust and TAM constructs such as usefulness and ease-of-use [30]; and along with trust, it influences network behaviour [37]. Further, it is not self-esteem or extroversion which predict successful online presence, but self-efficacy [72]. Indeed, as well as social pressure (see Social Norms above), self-efficacy affects the willingness to engage in online networks [68]. We therefore hypothesise that:

**H7a**: Computer self-efficacy is positively correlated with the behaviour in HMNs.

On the other hand, it appears that self-efficacy is negatively correlated with anxiety [73], which may be associated with perceived risk [74]. So a second hypothesis obtains:

**H7b**: Computer self-efficacy is negatively correlated with perceived risk.

## 4   Research Design

Having established an initial research model and formulated a set of hypotheses (see **Table 1**, below) based on our review of pertinent literature over the past decade, we are now in the process of organising both qualitative and quantitative investigation of that model. Following a similar qualitative approach to [75], we are starting with a focus group of those familiar with trust as a concept, how it is traditionally thought to relate to human-to-human interactions, and how it may transfer to technology – our expert group – to provide an initial evaluation of our research model. We are targeting six to ten participants for this group. Using the feedback from that group to identify potential refinements, we will then conduct a quantitative survey using a set of questions based on the instruments suggested by researchers in our literature review [30, 31, 37, 48], but extended and updated to reflect experience in the specific environment of HMNs [20, 48]. In an attempt to avoid the demographic constraints in many studies where respondents are confined to undergraduate students or a similar cohort, we are creating a publically available survey to be hosted by the University of Southampton which will run for approximately four weeks. We will combine this with snowball participant sampling if necessary to achieve a target of some 200 valid responses. These will be analysed using a structural-equation modelling analysis in line with work reported in [48]. On this basis, we hope to be in a position to report our results and present a validated research model associated with our hypotheses in the coming months.

**Table 1.** Hypotheses

| | |
|---|---|
| H1a | Perceived risk is negatively correlated with changes in regulation |
| H1b | Machine agency is negatively correlated with increasing levels of regulation |
| H1c | Regulation affects human agency |
| H2a | The perception of risk negatively affects trust in an HMN |
| H2b | The perception of risk negatively affects the behaviour in an HMN |
| H3a | HMN trust will positively influence the behaviour in an HMN |
| H3b | HMN trust will mediate human agency and the behaviour in an HMN |
| H4a | Social norms will affect the extent to which perceived risk influences behaviour in an HMN |
| H4b | Social norms will directly and positively correlate with behaviour in an HMN |
| H5a | Trust in the HMN is positively correlated with human agency |
| H5b | Human agency is related to self-efficacy |
| H6a | Machine agency in an HMN is directly related to human agency |
| H6b | Machine agency affects computer self-efficacy |
| H7a | Computer self-efficacy is positively correlated with the behaviour in HMNs |
| H7b | Computer self-efficacy is negatively correlated with perceived risk |

## 5 Conclusions and Future Work

Based on a review of current work on computer self-efficacy, agency and trust, we have developed a model which extends work reported by Thatcher et al [31] to include a set of constructs known to influence these constructs and behaviour within online networks. Validating this model will increase our understanding of online behaviours which is of interest to those who engage with the networks, but also those who seek either to monitor and understand or regulate online behaviours, as well as those building networks who wish to explore factors which will support the long-term health of that network. More especially, our model seeks to extend our understanding of the interplay between agency and trust on the one hand, but also self-efficacy and indeed social influence on the other. What we have proposed is, therefore, intended to advance our general understanding of interactions between human and machine agency in human-machine networks. In this way, we hope to throw some light on how conscious as well as programmed agency influence one another and affect the willingness to engage online as well as individual self-belief in the capability to achieve personal goals.

**Acknowledgements.** This work has been conducted as part of the HUMANE project, which has received funding from the European Union's Horizon 2020 research and innovation programme under grant agreement No 645043.